# Phase-field crystal modelling of crystal nucleation, heteroepitaxy and patterning

LÁSZLÓ GRÁNÁSY†‡*, GYÖRGY TEGZE†, GYULA I. TÓTH† and TAMÁS PUSZTAI†

†Research Institute for Solid State Physics and Optics, H-1525 Budapest,
POB 49, Hungary

‡Brunel Centre for Advanced Solidification Technology, Brunel University,
Uxbridge, Middlesex, UB8 3PH, UK

We apply a simple dynamical density functional theory, the phase-field-crystal (PFC) model, to describe homogeneous and heterogeneous crystal nucleation in 2d monodisperse colloidal systems and crystal nucleation in highly compressed Fe liquid. External periodic potentials are used to approximate inert crystalline substrates in addressing heterogeneous nucleation. In agreement with experiments in 2d colloids, the PFC model predicts that in 2d supersaturated liquids, crystalline freezing starts with homogeneous crystal nucleation without the occurrence of the hexatic phase. At extreme supersaturations crystal nucleation happens after the appearance of an amorphous precursor phase both in 2d and 3d. We demonstrate that contrary to expectations based on the classical nucleation theory, corners are not necessarily favourable places for crystal nucleation. Finally, we show that adding external potential terms to the free energy, the PFC theory can be used to model colloid patterning experiments.

Keywords: density functional theory, nucleation, heteroepitaxy, patterning, colloids, freezing in 2d

## 1. Introduction

The order of the liquid-solid transition depends on dimensionality. According to the theoretical expectations of Kosterlitz, Thouless, Halperin, Nelson, and Young (KTHNY) [1], in two dimensions (2d), melting takes place in two second-order transitions at two distinct temperatures ($T_m$ and $T_i$) [2]. First the dissociation of thermally activated dislocation pairs transforms the crystal into an orientationally ordered (hexatic) phase at the melting temperature $T_m$ and then the dissociation of free dislocations drives the system to form an isotropic fluid at $T_i > T_m$. This view is supported by computer simulations [3−6] and by experiments on colloidal systems [7−11], however, the order of the two transitions seems to depend on details of the inter-particle interaction and finite-size effects [12]. Some computer simulations indicate that the hexatic phase is metastable [13, 14]. Remarkably, experiments in 2d colloidal systems suggest that crystallization after deep quenching happens by direct nucleation of the crystalline phase from the liquid [15] implying a first-order transition. Interestingly, in some 2d colloidal systems a two-step crystallization process is observed, however, the precursor phase is rather an amorphous phase, not the hexatic phase [16−18].

In three dimensional non-equilibrium liquids, crystallization is a first-order phase transition, and the crystalline phase appears via nucleation, a process in which hetero-phase fluctuations form, whose atomic structure resembles to that of the crystalline phase at their central part [19, 20], while a continuous transition to the liquid phase is seen in the interfacial layer surrounding this central part [21]. While the intrinsic thickness of the interfacial layer might be fairly small [22], due to capillary waves the shape of these particles fluctuates, so that its time average may lead to a broader, diffuse interface as in the case of planar interfaces [22]. In qualitative agreement with the classical nucleation theory, the free energy of formation of these heterophase fluctuations shows a maximum as a function of size [19]. The maximum is the critical fluctuation (or nucleus), and represents a thermodynamic barrier: fluctuations that are larger grow with a high probability, while the smaller ones decay. It appears that in agreement with Ostwald's step rule, the first appearing solid is not necessarily the stable crystalline phase; it might be a metastable phase, whose atomic structure is closer to the structure of the liquid than the stable crystalline phase [23]. For example, there are theoretical expectations that in simple liquids the first nucleating phase has the bcc structure [24−26]. Indeed this expectation is supported by atomistic simulations for the Lennard-Jones system [21, 27] and by experiments showing metastable bcc nucleation in supersaturated superfluid $^4$He, in preference to the stable hcp phase [28]. Atomistic models based on the density functional technique (DFT) suggest that crystallization might happen via a dense liquid/amorphous precursor phase [29, 30] reminiscent to the two-step transition seen in 2d colloidal systems [16−18]. Other theoretical work implies that the presence of a metastable fluid critical point might assist crystal nucleation via a dense liquid precursor [31−35]. These findings raise the possibility that the two-step crystal nucleation via a precursor phase might be a fairly general phenomenon both in 2d and 3d. As for the structure of the (probably metastable) precursor, it may be amorphous or crystalline, depending on the system.

---
* Corresponding author. Email: grana@szfki.hu or Laszlo.Granasy@brunel.ac.uk

The formation of the hetero-phase fluctuations can be assisted by the presence of heterogeneities in the liquid, such as solid walls, floating solid particles, free surfaces, etc. Their main effect is that their atomic arrangement may induce ordering in the liquid adjacent to the wall [36−39]. This ordering of the liquid either helps or prevents the formation of heterophase fluctuations [40]. When the structure of the ordered liquid layer is compatible with the crystal structure to which the liquid freezes, the formation of heterophase fluctuations is enhanced at the wall, a phenomenon termed *heterogeneous nucleation* [41], as opposed to *homogeneous nucleation*, where the only heterogeneities are the internal fluctuations of the liquid phase. Heterogeneous nucleation is probably the most ubiquitous mechanism to start crystallization of undercooled liquids. It plays an essential role in determining the microstructure of crystalline materials, and has a continuously growing importance in manipulating crystallization morphology on the nanoscale [42−47]. Unfortunately, in practical cases little is known of liquid ordering and/or the molecular interaction between the wall and the solid and liquid phases. The heterogeneous nucleation process depends on atomistic details, such as the structure of the wall, its chemical properties, surface roughness, and ordering of the liquid at the wall, etc. The classical approach to heterogeneous nucleation relates the nucleation barrier to the *equilibrium contact angle* $\vartheta$, which in turn reflects the relative magnitudes of the wall-solid ($\gamma_{WS}$), wall-liquid ($\gamma_{WL}$), and solid-liquid ($\gamma_{SL}$) interfacial free energies [48]: $\cos\vartheta = (\gamma_{WS} - \gamma_{WL})/\gamma_{SL}$. It relies on the droplet or capillarity approximation that neglects the anisotropy of the interfacial free energies, and regards the interfaces as mathematically sharp. Then the critical fluctuation for homogeneous nucleation is spherical, has the radius $R^* = 2\gamma_{SL}/\Delta g$, while the nucleation barrier is $W_{hom} = (16\pi/3)(\gamma_{SL}^3/\Delta g^2)$, where $\Delta g$ is the driving force for solidification (the grand potential difference between the bulk solid and liquid phases). In this approximation, in the presence of a flat wall, only that fraction of the homogeneous nucleus (a spherical cap) needs to be created by thermal fluctuations, which realizes the contact angle at the triple junction line. Accordingly, in the heterogeneous case, the nucleation barrier is reduced by the catalytic potency factor $f(\vartheta) = (2+\cos\vartheta)(1-\cos\vartheta)^2/4 \leq 1$, so that $W_{het} = W_{hom} f(\vartheta)$ [49, 50]. Accordingly, 2d and 3d corners and conical cavities are preferred nucleation sites [50, 51]. While the classical model of heterogeneous nucleation captures some trends qualitatively [41, 51], it can be expected to be accurate for only large sizes. For example, in most cases, the nuclei are comparable in size to the interface thickness (as in [19, 21]), raising doubts concerning the applicability of the classical (sharp interface) droplet model. Indeed in the case of homogeneous nucleation in the hard-sphere system, the droplet model fails spectacularly [19]. An interesting and practically important limit, in which quantitative predictions are possible for foreign-particle-induced crystallization is, when these particles are ideally wet by the crystalline phase, i.e., the nucleation process is avoided and the conditions of free growth limit the ability of a particle to start crystallization. This case has been investigated extensively by Greer and co-workers [52−54].

Some of the difficulties associated with the classical model of heterogeneous crystal nucleation can be removed using advanced continuum models such as the phase-field theory [55−61]; non-classical effects including liquid ordering at the walls [59, 60], the presence of surface spinodals [59, 60] or nucleation of an intermediate phase on the substrate [61] can also be addressed. Unfortunately, usually it is difficult to relate the parameters of these coarse-grained models to microscopic features.

In order to handle the interaction between the substrate and the solidifying liquid one needs an atomistic approach. Atomistic simulations, such as molecular dynamics and Monte Carlo simulations have provided important information on the microscopic aspects of the wetting of foreign walls by liquid and crystal [36–39, 62]. Recent Monte Carlo studies revealed the importance of the line tension [36, 62] in the case of unstructured walls. It has been also shown that for large clusters the classical description of heterogeneous nucleation works well, while deviations are observed at large undercoolings / supersaturations [62]. Furthermore, the interfacial free energies for flat and *curved* interfaces have also been evaluated [62]. Other atomistic techniques such as the dynamical density functional (DDFT) theory have been used to address the effect of crystalline seeds of tuneable structure on the process of crystallization [63]. A recently developed simple DDFT-type approach, termed the phase-field crystal (PFC) model [64], has been used to investigate heterogeneous nucleation on unstructured walls [65]. 2d PFC simulations have also been used to explore pattern formation on periodic substrates represented by periodic potentials [66]. Despite these advances, further atomistic studies of the effect of patterned/crystalline substrates on crystal nucleation and pattern formation are warranted.

In the present work, we use the PFC theory to model homogeneous and heterogeneous crystal nucleation in 2d and 3d and to describe colloidal pattern formation in 2d. We concentrate here on the following issues: (i) Appearance of a precursor phase in homogeneous nucleation; (ii) heterogeneous nucleation on crystalline substrates; and (iii) modelling of 2d colloidal patterning experiments. In the case of heterogeneous nucleation and pattern formation, the PFC model is supplemented with an appropriate potential energy term. To study homogeneous crystal nucleation in 3d metallic materials and possible appearance of precursors, we adopt a phenomenological extension of the PFC, which is able to reproduce the interfacial properties of bcc Fe fairly well [67].

## 2. The phase-field crystal model (PFC)

The PFC model is a simple DDFT type approach introduced by Elder and co-workers [64, 68]. Its free energy functional can be deduced [68] from the Ramakrishnan-Yussouff type perturbative DFT [69] after some simplifications that lead to a Brazowskii/Swift-Hohenberg form [70, 71], while the time evolution is governed by an overdamped conservative equation of motion [64, 68]. The relationship between the DDFT and PFC has been further clarified in [72]. The PFC model has been used successfully to address elasticity and grain boundaries [68], the anisotropies of the interfacial free energy [73, 74] and growth rate [75], dendritic and eutectic growth [76–79], glass formation [30], melting at dislocations and grain boundaries [80, 81], and polymorphism [75]. While it is a microscopic approach, it has the advantage over other classical microscopic techniques, such as molecular dynamics simulations that the time evolution of the system can be studied on the many orders of magnitude longer diffusive time scale, so that the long-time behaviour and the large-scale structures become accessible. We note that the diffusion-controlled relaxation dynamics the PFC model assumes is especially relevant for colloidal systems [63, 72], where the self-diffusion of the particles is expected to be the dominant way of density relaxation. For normal liquids at small undercoolings, the hydrodynamic mode of density relaxation dominates, which might be approximately incorporated by adding a term proportional to $\partial^2 n/\partial t^2$ [82].

### 2.1. *Free energy functional*

Following Elder and Grant [68], in deriving the free energy functional of the PFC, we start from the perturbative density functional theory by Ramakrishnan and Yussouff [69], in which the free energy difference $\Delta F = F - F_L^{ref}$ between the crystal and a reference liquid of particle density $\rho_L^{ref}$ can be written in the following form, after truncating the Taylor expansion above the two-particle term:

$$\frac{\Delta F}{kT} = \int d\mathbf{r} \left[ \rho \ln\left(\frac{\rho}{\rho_L^{ref}}\right) - \Delta\rho \right] - \frac{1}{2} \iint d\mathbf{r}_1 d\mathbf{r}_2 [\Delta\rho(\mathbf{r}_1) C(\mathbf{r}_1,\mathbf{r}_2) \Delta\rho(\mathbf{r}_2)] + ... \qquad (1)$$

where $\Delta\rho = \rho - \rho_L^{ref}$, while $C(\mathbf{r}_1,\mathbf{r}_2)$ is the two-particle direct correlation function of the reference liquid. The density of the solid can be Fourier expanded as $\rho_S = \rho_L^{ref} \{1 + \eta_S + \sum_\mathbf{K} A_\mathbf{K} \cdot \exp(i\mathbf{Kr})\}$, where $\eta_S$ is the fractional density change between the solid and the reference liquid, while **K** are reciprocal lattice vectors, and $A_\mathbf{K}$ are the respective Fourier amplitudes. Introducing the reduced number density relative to the reference liquid, $n = (\rho - \rho_L^{ref})/\rho_L^{ref}$, one finds that $\rho = (1 + n) \rho_L^{ref}$, while $n = \eta_S + \sum_\mathbf{K} A_\mathbf{K} \cdot \exp(i\mathbf{Kr})$; thus

$$\frac{\Delta F}{kT} = \int d\mathbf{r}\left[\rho_L^{ref}(1+n)\ln(1+n) - \rho_L^{ref} n\right] - \frac{1}{2}\iint d\mathbf{r}_1 d\mathbf{r}_2 \left[\rho_L^{ref} n(\mathbf{r}_1) C(\mathbf{r}_1,\mathbf{r}_2) \rho_L^{ref} n(\mathbf{r}_2)\right] + \ldots \quad (2)$$

To arrive to the free energy functional used in the PFC, we expand $C(\mathbf{r}_1,\mathbf{r}_2)$ in Fourier space, $\hat{C}(k) \approx \hat{C}_0 + \hat{C}_2 k^2 + \hat{C}_4 k^4 + \ldots$, where $\hat{C}(k)$ has its 1st peak at $k = 2\pi/\sigma$, while the sign of the coefficients is expected to alternate and $\sigma$ is the inter-particle distance. We introduce the dimensionless form of $\hat{C}(k)$ as $c(k) = \rho_L^{ref} \hat{C}(k) \approx \sum_{j=0}^{m} c_{2j} k^{2j} = \sum_{j=0}^{m} b_{2j}(k\sigma)^{2j}$, which is related to the structure factor as $S(k) = 1/[1 - c(k)]$. Returning to real space, the free energy difference reads as

$$\frac{\Delta F}{kT\rho_L^{ref}} \approx \int d\mathbf{r}[(1+n)\ln(1+n) - n] - \frac{1}{2}\iint d\mathbf{r}_1 d\mathbf{r}_2 \left[ n(\mathbf{r}_1)\left\{\sum_{j=0}^{m}(-1)^j c_{2j}\nabla^{2j}\right\}\delta(\mathbf{r}_1 - \mathbf{r}_2)n(\mathbf{r}_2) \right]. \quad (3)$$

After integrating the second term on RHS with respect to $\mathbf{r}_2$ and replacing $\mathbf{r}_1$ by $\mathbf{r}$, we find

$$\frac{\Delta F}{kT\rho_L^{ref}} \approx \int d\mathbf{r}\left[(1+n)\ln(1+n) - n - \frac{n}{2}\left\{\sum_{j=0}^{m}(-1)^j c_{2j}\nabla^{2j}\right\}n\right]. \quad (4)$$

Note that the reference liquid (of particle density $\rho_L^{ref}$) is not necessarily the initial liquid. The particle density of the latter we denote as $\rho_L^0$. For this initial liquid, the reduced density, $n_L^0 = (\rho_L^0 - \rho_L^{ref})/\rho_L^{ref}$, may differ from 0. Accordingly, this initial liquid state might be considered as a liquid either compressed or stretched relative to the reference liquid. As a result, we may have now two parameters to control the driving force for solidification: the initial liquid number density $n_L^0$ (not far from the reference), and the temperature, if the direct correlation function depends on temperature. Taylor-expanding $\ln(1 + n)$ for small $n$:

$$\ln(1 + n) \approx n - n^2/2 + n^3/3 - n^4/4 + \ldots \qquad \text{for } |n| < 1$$

thus

$$(1 + n)\ln(1 + n) \approx n + n^2/2 - n^3/6 + n^4/12 - \ldots$$

and finally one obtains

$$\frac{\Delta F}{kT\rho_L^{ref}} \approx \int d\mathbf{r}\left[\frac{n^2}{2} - \frac{n^3}{6} + \frac{n^4}{12} - \frac{n}{2}\left\{\sum_{j=0}^{m}(-1)^j c_{2j}\nabla^{2j}\right\}n\right]. \quad (5)$$

For the $m = 2$ used in the simplest version of PFC [64, 68] and considering the alternating sign of the expansion coefficients of $\hat{C}(k)$, Eq. (5) boils down to the following form:

$$\Delta F \approx kT\rho_L^{ref} \int d\mathbf{r}\left\{\frac{n^2}{2}(1+|b_0|) + \frac{n}{2}\left[|b_2|\sigma^2\nabla^2 + |b_4|\sigma^4\nabla^4\right]n - \frac{n^3}{6} + \frac{n^4}{12}\right\}. \quad (6)$$

Introducing the new variables

$B_L = 1 + |b_0| = 1 - c_0$     [$= (1/\kappa)/(\rho_L^{ref} kT)$, where $\kappa$ is the compressibility],

$B_S = |b_2|^2/(4|b_4|)$     [$= K/(\rho_L^{ref} kT)$, where $K$ is the bulk modulus of the crystal],

$R = \sigma(2|b_4|/|b_2|)^{1/2}$     [$=$ the new length scale ($x = R \cdot \tilde{x}$), which is now related to the position of the maximum of the Taylor expanded $\hat{C}(k)$], and a multiplier $v$ for the $n^3$ term ($v$ accounts here for the 3-particle correlation in 0th order), one obtains an equation analogous to the one used by Berry *at al.* [30] in their paper on glass transition:

$$\Delta F = \int d\mathbf{r} I(n) = kT\rho_L^{ref} \int d\mathbf{r}\left\{\frac{n}{2}\left[B_L + B_S(2R^2\nabla^2 + R^4\nabla^4)\right]n - v\frac{n^3}{6} + \frac{n^4}{12}\right\}, \quad (7)$$

where $I$ is the total (dimensional) free energy density.

*Conversion to the Swift-Hohenberg formalism:* Introducing the new variables $x = R \cdot \tilde{x}$, $n = (3B_S)^{1/2}\psi$, $\Delta F = (3\rho_L^{ref} kTR^d B_S^2) \cdot \Delta\tilde{F}$, the free energy functional transforms into a modified Swift-Hohenberg-type dimensionless free energy:

$$\Delta\tilde{F} = \int d\tilde{\mathbf{r}}\left\{\frac{\psi}{2}\left[r^* + (1+\tilde{\nabla}^2)^2\right]\psi + t^*\frac{\psi^3}{3} + \frac{\psi^4}{4}\right\}, \quad (8)$$

where, $t^* = -(v/2)\cdot(3/B_S)^{1/2} = -v\cdot(3|b_4|/|b_2|^2)^{1/2}$ and $r^* = \Delta B/B_S = (1 + |b_0|)/[|b_2|^2/(4|b_4|)] - 1$, while $\psi = n/(3B_S)^{1/2}$. Note that all quantities involved in equation (8) including those with tilde are dimensionless. The form of equation (8) suggests that the free energy functional of the $m = 2$ PFC model contains only two dimensionless similarity parameters $r^*$ and $t^*$ that can be obtained as combinations of the original (physical) model parameters. We note, finally, that even the third-order term can be eliminated. In the respective $t^{*'} = 0$ Swift-Hohenberg model, the state $[r^{*'} = r^* - (t^*)^2/3, \psi' = \psi - t^*/3]$ corresponds to the state ($r^*$, $\psi$) of the original $t^* \neq 0$ model. This latter transformation leaves the grand canonical potential difference, the Euler-Lagrange equation and the equation of motion invariant. Therefore, it is sufficient to address the $t^* = 0$ case, as we do in the rest of this work.

*Eight-order fitting of C(k) (PFC EOF):* In a recent paper, Jaatinen *et al.* [67] have proposed an eight-order expansion of the two-particle direct correlation function in the Fourier space, however, now around its maximum ($k = k_m$):

$$C(k) \approx C(k_m) - \Gamma\left(\frac{k_m^2 - k^2}{k_m^2}\right)^2 - E_B\left(\frac{k_m^2 - k^2}{k_m^2}\right)^4, \quad (9)$$

where the expansion parameters were fixed so that the quantities $C(k = 0)$, $k_m$, $S(k_m)$ and $C''(k_m)$, i.e., the liquid compressibility and the position, height, and the second derivative of $C(k)$ are accurately recovered. This is ensured by

$$\Gamma = -\frac{k_m^2 C''(k_m)}{8} \quad \text{and} \quad E_B = C(k_m) - C(0) - \Gamma. \quad (10)$$

This choice of parameters with input data for Fe from [73] has led to a fair agreement with molecular dynamics results for the volume change upon melting, the bulk modulus of the liquid and solid phases, and the magnitude and anisotropy of the solid-liquid interfacial free energy [67].

## 2.2. *The equation of motion*

In analogy to the DDFT [63, 72], we assume conserved dynamics, however, with a mobility coefficient of $M_\rho = \rho_0 D_\rho/kT$. The respective (dimensional) equation of motion is

$$\frac{\partial\rho}{\partial t} = \nabla\left\{M_\rho\left[\nabla\frac{\delta\Delta F}{\delta\rho}\right] + \left(\frac{2kTM_\rho}{\Delta x^d \Delta t}\right)^{1/2}\mathcal{N}\right\} = \nabla\left\{\left(\frac{\rho_0 D_\rho}{kT}\right)\nabla\left[\frac{\partial I}{\partial\rho} + \sum_j (-1)^j \nabla^j \frac{\partial I}{\partial\nabla^j\rho}\right] + \left(\frac{2\rho_0 D_\rho}{\Delta x^d \Delta t}\right)^{1/2}\mathcal{N}\right\}, \quad (11)$$

where the second term in the central expression represents the discretized form of a conserved fluctuation-dissipation noise [83], while $\mathcal{N}$ is a Gaussian white noise of standard deviation 1. To obtain a dimensionless form, first we change from variable $\rho$ to $n$, yielding

$$\frac{\partial n}{\partial t} = \nabla\left\{\left(\frac{(1+n_0)D_\rho}{kT\rho_L^{ref}}\right)\nabla\left[\frac{\partial I}{\partial n} + \sum_j (-1)^j \nabla^j \frac{\partial I}{\partial\nabla^j n}\right] + \left(\frac{2kT(1+n_0)D_\rho}{kT\rho_L^{ref}\Delta x^d \Delta t}\right)^{1/2}\mathcal{N}\right\} = \nabla\left\{M_n\nabla\left[\frac{\delta\Delta F}{\delta n}\right] + \left(\frac{2kTM_n}{\Delta x^d \Delta t}\right)^{1/2}\mathcal{N}\right\}$$

where we have introduced $M_n = [(1+n_0) D_\rho /(kT\rho_L^{ref})]$. Scaling the time and distance as $t = \tau \cdot \tilde{t}$ and $x = \sigma \cdot \tilde{x}$, where $\tau = \sigma^2/[D_\rho (1+ n_0)]$, we find

$$\frac{\partial n}{\partial \tilde{t}} = \tilde{\nabla}\left\{\left(\frac{1}{kT\rho_L^{ref}}\right)\tilde{\nabla}\left[\frac{\partial I}{\partial n}+\sum_j (-1)^j \nabla^j \frac{\partial I}{\partial \nabla^j n}\right]+\left(\frac{2}{\rho_L^{ref}\sigma^d \Delta \tilde{x}^d \Delta \tilde{t}}\right)^{1/2}\mathcal{N}\right\},$$

where $I$ is the total free energy density defined above.

Inserting $I$ from equation (6), we obtain

$$\frac{\partial I}{\partial n} = kT\rho_L^{ref}\left\{n(1+|b_0|)+\frac{1}{2}\sum_{j=1}^m |b_{2j}|\tilde{\nabla}^{2j}n-\frac{n^2}{2}+\frac{n^3}{3}\right\},$$

$$\tilde{\nabla}^j \frac{\partial I}{\partial \tilde{\nabla}^j n} = \tilde{\nabla}^j\left(\frac{n}{2}kT\rho_L^{ref}|b_j|\right) = \frac{1}{2}kT\rho_L^{ref}|b_j|(\tilde{\nabla}^j n) \qquad j > 0$$

leading to the following dimensionless equation of motion:

$$\frac{\partial n}{\partial \tilde{t}} = \tilde{\nabla}^2\left[n(1+|b_0|)+\sum_{j=1}^m |b_{2j}|\tilde{\nabla}^{2j}n-\frac{n^2}{2}+\frac{n^3}{3}\right]+\tilde{\nabla}\left(\frac{2}{\rho_L^{ref}\sigma^d \Delta \tilde{x}^d \Delta \tilde{t}}\right)^{1/2}\mathcal{N}. \qquad (12)$$

Analogously, if we start from equation (7), the equation of motion can be obtained as follows.

$$\frac{\partial n}{\partial t} = \nabla\left\{M_n \nabla\left[\frac{\delta \Delta F}{\delta n}\right]+\left(\frac{2kTM_n}{\Delta x^d \Delta t}\right)^{1/2}\mathcal{N}\right\} = \nabla\left\{M_n \nabla\left[\frac{\partial I}{\partial n}+\sum_j (-1)^j \nabla^j \frac{\partial I}{\partial \nabla^j n}\right]+\left(\frac{2kTM_n}{\Delta x^d \Delta t}\right)^{1/2}\mathcal{N}\right\} =$$

$$= \nabla\left\{M_n \nabla\left[(kT\rho_L^{ref})\left[B_L+B_S(2R^2\nabla^2+R^4\nabla^4)\right]n-v\frac{n^2}{2}+\frac{n^3}{3}\right]+\left(\frac{2kTM_n}{\Delta x^d \Delta t}\right)^{1/2}\mathcal{N}\right\} \qquad (13)$$

where $M_n = [(1+n_0)D_\rho/(kT\rho_L^{ref})]$.

*Equation of motion in the Swift-Hohenberg type dimensionless formalism:* Introducing the new variables $t = \tau \cdot \tilde{t}$, $x = R \cdot \tilde{x}$, and $n = (3B_S)^{1/2}\psi = (3B_S)^{1/2}[\psi' + t^*/3]$ into equation (13), where $\tau = R^2/(B_S M_n \rho_L^{ref} kT)$, one arrives to the equation of motion of the earliest PFC model [64]

$$\frac{\partial \psi'}{\partial \tilde{t}} = \tilde{\nabla}^2\left\{[r^{*\prime}+(1+\tilde{\nabla}^2)^2]\psi'+\psi'^3\right\}+\tilde{\nabla}\left(\frac{\alpha^*}{\Delta \tilde{x}^d \Delta \tilde{t}}\right)^{1/2}\mathcal{N}, \qquad (14)$$

where $r^{*\prime} = r^* - (t^*)^2/3 = [\Delta B - (v/2)^2]/B_S = (1 + |b_0|)/[|b_2|^2/(4|b_4|)] - [1 + v^2\cdot(|b_4|/|b_2|^2)]$ and the dimensionless noise strength is $\alpha^* = 2/(3B_S^2 \rho_L^{ref} R^d) = 2^{5-d/2}|b_4|^{2-d/2}/[3\sigma^d \rho_L^{ref}|b_2|^{4-d/2}]$, while the correlator for the dimensionless noise reads as $\langle \zeta(\tilde{\mathbf{r}},\tilde{t}),\zeta(\tilde{\mathbf{r}}',\tilde{t}')\rangle = \alpha^*\cdot \tilde{\nabla}^2 \delta(\tilde{\mathbf{r}} - \tilde{\mathbf{r}}')\cdot \delta(\tilde{t} - \tilde{t}')$. Summarizing, the dynamical $m = 2$ PFC model has two dimensionless similarity parameters $r^{*\prime}$ and $\alpha^*$ composed of the original (physical) model parameters.

The equation of motion has been solved numerically on uniform rectangular 2d grids using a fully spectral semi-implicit scheme described in [79] and periodic boundary condition at the perimeters. A parallel C code relying on the MPI protocol has been developed. To optimize the performance, we have developed a parallel FFT code based on the FFTW3 library [84]. The numerical simulations presented in this paper have been performed on two PC clusters: One at the Research Institute for Solid State Physics and Optics (RISSPO), Budapest, Hungary, that consists of 24 PCs, each equipped with two 2.33 GHz Intel processors of 4 CPU cores (192 CPU cores in all on the 24 nodes), 8 GB memory/node, and with 10 Gbit/s (InfiniBand) inter-node communication, and another cluster hosted by the Brunel Centre for Advanced Solidification Technology (BCAST), Brunel University, West London, UK, which consists of 20 similar nodes (160 CPU cores), however, with 1 Gbit/s (standard GigaBit Ethernet) communication in between.

Owing to the ensemble averaging inherent in the DFT type models, one is in principle unable to simulate crystal nucleation. This difficulty can be partly removed by adding Langevin noise (of a correlator that satisfies the fluctuation-dissipation theorem) to the equation of motion (see equations (11)–(14)) to represent thermal fluctuations. Unfortunately, this is not without conceptual difficulties (see the discussion in [85−87]): if the number density is considered as an ensemble averaged quantity, the fluctuations are already incorporated into the free energy functional, and the addition of noise will lead to double counting part of the fluctuations [85, 86]. If, however, the number density is regarded as a quantity coarse-grained in time, there is phenomenological motivation to add the noise [87]. The qualitative picture with noise is appealing: we see how nucleation and growth happen on the atomistic level therefore in this work we incorporate Langevin noise into the equation of motion. To avoid part of the associated difficulties, we have applied a coloured noise obtained by filtering out the unphysical short wavelengths that are smaller than the inter-particle distance (this also removes the ultraviolet catastrophe, expected in 3d [88], and the associated dependence of the results on spatial resolution). Some related issues and their solution via parameter renormalization [89] are addressed in another paper of this volume [90].

### 2.3. *Modelling of an inert substrate*

Preliminary simulations have shown that in the case of potentials that do not cover the whole simulation area (i.e., substrate and liquid appears together at the beginning), filling/depletion of the area covered by the potential leads to a transient, which establishes depletion/excess zone at their perimeter that relaxes extremely slowly due to the diffusional relaxation dynamics present in the PFC model, a relaxation that might interfere with crystallization. To reduce this effect, it makes sense to assign such a density field to the substrate that establishes equilibrium with the initial solid. To approximate this situation we use the single-mode approximation to find the equilibrium. In the case of an external potential of form $V(x, y) = V_0 + V_1 [\cos(qx) + \cos(qy)]$, where $q = 2\pi/a$ and $a$ is the lattice constant of the external potential, we have used the *ansatz* $\psi(x, y) = \psi_S + A [\cos(qx) + \cos(qy)]$ to find that density distribution, which is in equilibrium with the liquid phase of reduced particle density $\psi_0$. This ansatz is a reasonable approximation if the potential is strong enough to dominate over the natural 2d hexagonal structure the PFC realizes in 2d. In practice, for a set of fixed $\psi_0$, $a$, $V_1$, and $r^*$, we have adjusted $V_0$ until the respective substrate solution characterized by $\psi_S$ and $A$ (found analytically) had a common tangent with the initial liquid phase. Starting the simulations with this initial density distribution in the substrate area has reduced significantly the initial transient. This technique has been applied in some of our simulations for heterogeneous nucleation in 2d.

### 2.4. *Parameters used in modelling*

Unless stated otherwise, in the 2d colloidal simulations, we have used the following parameters: $r^* = -0.75$, $\alpha^* = 0.1$, $\psi_0 = -0.5$, $\Delta\tau = 0.025$ and $\Delta x = \pi/8$. In the 3d PFC EOF simulations for Fe, we have used the same physical properties as in [67], however, we have increased the pressure (density) to drive the system out of equilibrium enough to initiate homogeneous nucleation.

### 3. Results and discussion

In this section, first we review a few relevant properties of the 2d PFC model, including its phase diagram and the anisotropy of the interfacial free energy as reflected by the equilibrium crystal shape. The subsequent sub-sections address first the homogeneous nucleation, followed by heteroepitaxy and heterogeneous nucleation. Finally, we present a few representative simulations for colloidal controlled colloidal self-assembly in 2d.

### 3.1. *Phase diagram and properties for 2d*

In 2D, the PFC model predicts the following stable phases [64, 68]: a homogenous disordered (fluid) phase, an ordered hexagonal (crystalline) phase, and a striped phase (see figure 1). An analogous model, based on the Brazowskii/Swift-Hohenberg type free energy, has been used widely to understand morphological transitions in block-copolymers, where these ordered and disordered phases have familiar realizations [91, 92]. In the strong coupling regime of the PFC model, where the magnitude of $r^*$ is large, a section of the fluid-crystal coexistence can be rescaled to real crystal-liquid systems as pointed out in [68]. Such rescaling can be utilized to define the reference liquid (whose particle density is $\rho_L^{ref}$) and the relationship between $r^*$ and the physical temperature. We note here that there is no convincing theoretical or experimental evidence for the existence of a critical/spinodal point between the crystalline and liquid phases in simple single component system [93–95]. Remarkably, however, a recent molecular dynamics study with a pair potential resembling to a Derjaguin-Landau-Verwey-Overbeek potential with a secondary minimum (often used for colloids) indicates the presence of a critical point between the solid and liquid phases [96].

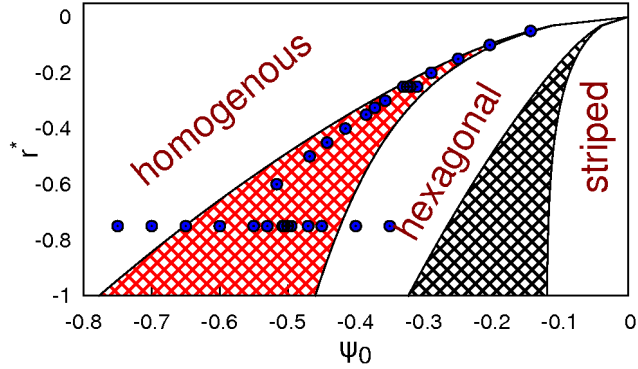

Figure 1. The $\psi < 0$ section of the PFC phase diagram predicted using the single-mode approximation [64]. The circles denote points in which simulations have been performed. Note the critical point at $\psi_c = 0$ and $r^*_c = 0$.

The state points at which our simulations have been performed and the relevant phase boundaries are indicated in figure 1.

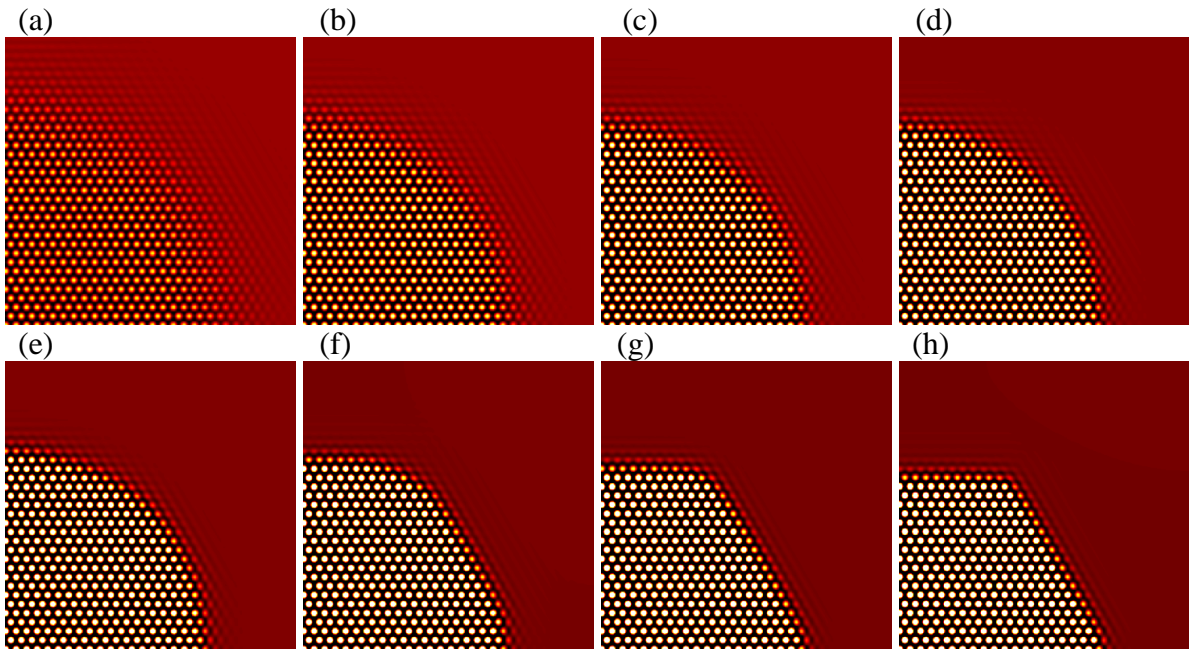

Figure 2. Equilibrium shape vs reduced temperature $r^*$ as predicted for a crystalline fraction of $X \sim 0.3$ in the absence of noise ($\alpha^* = 0$). (a) – (h): $r^* = -0.05, -0.10, -0.15, -0.20, -0.25, -0.30, -0.325$, and $-0.35$ (see the respective points in figure 1). Note that the interface thickness decreases while the anisotropy increases with an increasing distance from the critical point. The computations have been performed on a $1024 \times 1024$ rectangular grid (the upper right quarter of the simulations is shown). Equilibration has been performed for a period of $10^6$ dimensionless time steps. Reduced particle density maps are shown.

The equilibrium (Wulff) shapes have been determined in the reduced temperature range of $-0.75 \leq r^* \leq -0.05$ in the absence of noise ($\alpha^* = 0$). It has been obtained by placing a seven-particle cluster into the simulation box and letting it grow until establishing equilibrium. (For $r^* < -0.75$ accession of equilibrium becomes computationally prohibitive.) The initial liquid density has been chosen so that the expected crystalline fraction obtained from the lever rule is $X = 0.3$. Representative results are shown in figure 2. We observe a circular shape above $r^* \leq -0.25$, faceting for $r^* \leq -0.25$, and a hexagonal shape with sharp corners below $r^* \leq -0.325$. As expected from atomistic simulations and various theoretical treatments [97], the interface thickness diverges at the critical point in figure 1, while the anisotropy increases with increasing distance from the critical point. It has also been observed that the amplitude of the density peaks representing the atoms increases with decreasing $r^*$. The present results are in general agreement with those of a recent work [98] however our investigations have been performed in a somewhat broader range of the reduced temperature. In 2d, it is expected that the solid-liquid interface roughens for any non-zero temperature [97, 99, 100]. For example, in the case of the 2d Ising model on a triangular lattice, a hexagonal Wulff shape is obtained at $T = 0$, whose corners become increasingly rounded with increasing temperature until the equilibrium shape becomes a circle at the critical point $T = T_c$ [99]. One may wish to scale the PFC results for the equilibrium shape onto exact results for the 2d Ising model, to obtain an approximate temperature scale for the PFC. However, this should be done for PFC equilibrium shapes obtained in the presence of noise ($\alpha^* > 0$) to ensure that the two models belong to the same universality class.

To model the strongly faceted nature of the 2d colloidal crystal aggregates (evident in many of the experimental images [101, 102]), which implies that the anisotropy is beyond the limit that leads to excluded orientations, we have chosen here $r^* = -0.75$ for our simulations for colloids (the corresponding reduced equilibrium densities for the liquid and solid phases are $\psi_L^e = -0.6514$ and $\psi_{Hex}^e = -0.4228$, respectively). In a few cases, we wished to investigate systems that are closer to metals, therefore, we have chosen $r^* = -0.25$ that corresponds to a system lying on the border of faceting (in this case $\psi_L^e = -0.3388$ and $\psi_{Hex}^e = -0.2907$).

### 3.2. *Modelling of homogeneous nucleation and growth in 2d*

First, we investigate the nucleation pathways as a function of supersaturation with a choice of the PFC model parameters ($r^* = -0.75$ and varied $\alpha^*$) that reproduce fairly the faceting seen in 2d colloidal systems. Our main interest here is whether the formation of a precursor phase might indeed be observed. Candidates that might appear are the amorphous [16–18] and hexatic [1] phases. In order to characterize the structural features of the solid matter, we have determined the pair-correlation function and the bond-order correlation function. The latter is defined as

$$g_6(r) = \frac{\langle \Psi_6^*(\mathbf{r}')\Psi_6(\mathbf{r}'-\mathbf{r}) \rangle}{\langle \delta(\mathbf{r}')\delta(\mathbf{r}'-\mathbf{r}) \rangle} = \frac{\sum_i \sum_{i<j} \Psi_6^*(\mathbf{r}_i)\Psi_6(\mathbf{r}_i - \mathbf{r}_j)}{g(r)}, \quad (15)$$

where $g(r) = \langle \delta(\mathbf{r}')\delta(\mathbf{r}' - \mathbf{r}) \rangle$ is the pair correlation function and

$$\Psi_6(\mathbf{r}_l) = \frac{1}{n_l}\sum_{j=1}^{n_l} e^{i6\theta_{lj}},$$

while the summation is performed for the $n_l$ nearest neighbours of particle $l$ positioned at $\mathbf{r}_l$, $\theta_{lj}$ is the angle between a fixed reference orientation and the bond between particle $l$ and its neighbour $j$, while the averaging is performed for all the $N(N-1)/2$ particle-pair distances [1, 5, 11]. According to the KTHNY theory

$$\lim_{r\to\infty} g_6(r) \neq 0 \quad \text{crystal: long-range order;}$$
$$g_6(r) \propto r^{-\eta_6} \quad \text{hexatic: quasi-long-range order;}$$
$$g_6(r) \propto e^{-r/\xi_6} \quad \text{isotropic: short-range order.}$$

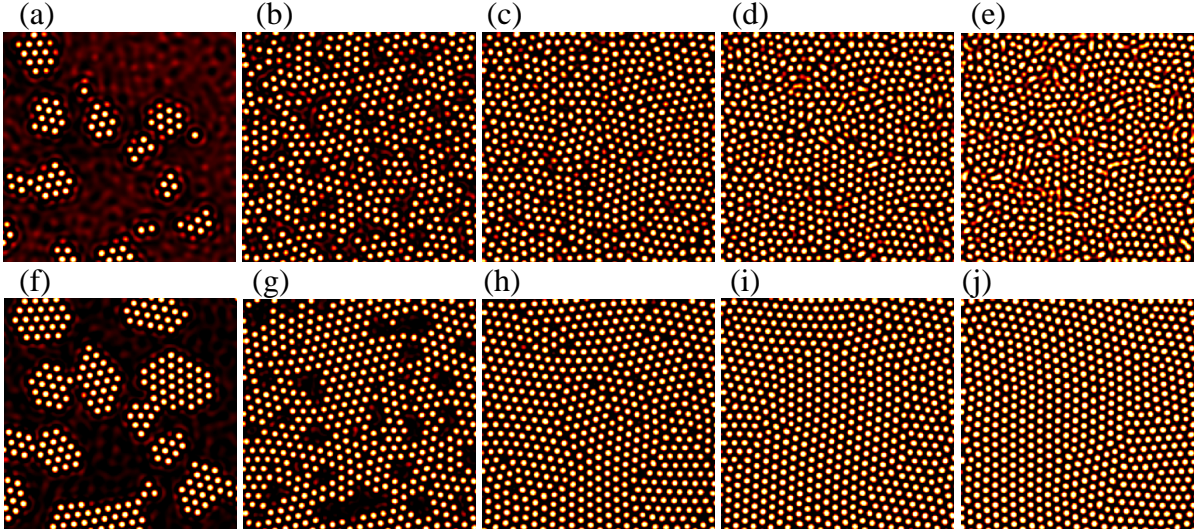

Figure 3. Snapshots of early and late stages of solidification in PFC simulations performed with initial reduced particle densities of $\psi_0 = -0.55, -0.50, -0.45, -0.40$ and $-0.35$. (a)–(e) Early stage: The respective reduced times are: $\tau/\Delta\tau = 10000, 3000, 1500, 1000,$ and $700$. (f)–(j) Late stage: The same areas are shown at reduced time $\tau/\Delta\tau = 60000$. ($418 \times 418$ fractions of $2048 \times 2048$ sized simulations are shown. Other simulation parameters were: $r^* = -0.75$ and $\alpha^* = 0.1$. Reduced particle density maps are shown.)

The exponent $\eta_6$, that describes the spatial decay of the bond-order correlation function, is smaller than ¼ in the hexatic phase, and is ¼ at $T = T_i$, while $\xi_6$ is the bond-order correlation length. Thus, investigation of the bond-order correlation function can help to identify the phases one observes during freezing [1, 5, 11, 15].

We have performed a set of simulations on a $2048 \times 2048$ grid with noise strength of $\alpha^* = 0.1$, while varying the initial particle density as $\psi_0 = -0.55, -0.50, -0.45, -0.40$ and $-0.35$. The early stage and late stage morphologies are shown in figure 3. While at low initial particle densities individual crystallites nucleate directly from the liquid, with increasing initial density solidification happens fairly simultaneously everywhere and results in an increasingly disordered structure with increasing $\psi_0$. Whether the latter should be viewed as highly polycrystalline matter of extremely small grain size or an amorphous phase is not immediately clear.

The respective pair correlation functions $g(r)$ and bond-order correlation functions $g_6(r)$ are shown in figures 4–5. The pair correlation functions corresponding to the structures shown in figures 3c–e appear to be similar to those expected for liquid/amorphous phases [103–105]. For comparison, we have also evaluated $g(r)$ for the clearly microcrystalline structure shown in figure 3b ($\psi_0 = -0.50$).

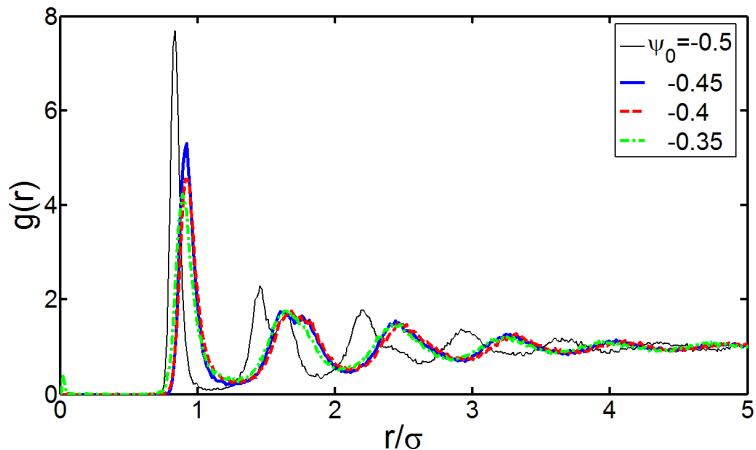

Figure 4. Pair correlation function for the early stage solidification structures shown in figures 3b–e.

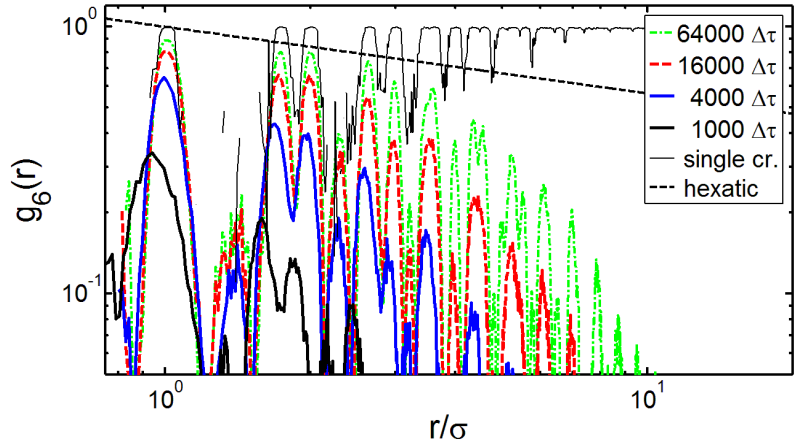

Figure 5. Time evolution of the bond-order correlation function for $\psi_0 = -0.40$ on log-log scale. $g_6(r)$ is shown at $\tau/\Delta\tau = 1000, 4000, 16\,000$, and $64\,000$. For comparison, the upper envelop expected for the hexatic phase and the result for a single crystal are also shown. These curves describe an amorphous to polycrystalline transition (cf. figures 3d and 3i). Note that the upper envelope of the $g_6(r)$ curves decay faster than expected for the hexatic phase.

As expected, here the peaks are sharper, the second peak is split, and for $r/\sigma > 3$ the peaks are in antiphase relative to those for figures 3c–e. Indeed, even to the naked eye, the structures shown in figures 3c–e are considerably more disordered than the microcrystalline structure displayed in figure 3b. We also note that their degree of disorder appears to be comparable to that of the experimental liquids shown in figures 4b and 4c of [15]. We conclude thus that at sufficiently large particle densities first a disordered solid phase forms that consists of particles localized on the time scale of the simulation. As a result of its structural properties, this highly disordered solid phase is termed here as amorphous. We cannot, however, fully rule out that it has been formed via copious nucleation. With increasing time, in all these simulations a polycrystalline late stage has been achieved (see figure 3f–j).

The time evolution of the bond-order correlation function, we obtained for $\psi_0 = -0.40$, closely follows the behaviour seen in experiments [15], and clearly rules out the presence of a hexatic phase (figure 5): At no stage of freezing can one observe that the upper envelope of $g_6(r)$ is linear with a slope of $-1/4$ or less, as expected for the hexatic phase on log-log scale. Similar results have been obtained for the other two reduced particle densities ($\psi_0 = -0.45$ and $-0.35$).

Summarizing, at low thermodynamic driving forces the crystalline phase nucleates directly from the non-equilibrium liquid (as observed in some of the colloid experiments [15]), while at large driving forces an amorphous precursor phase appears (as has been seen in other colloidal systems [16–18]). Further work is needed, however, to clarify whether the present model is able to reproduce crystal nucleation inside amorphous particle rafts floating in the liquid phase, as observed in [16].

### 3.3. Modelling of homogeneous nucleation in 3d

To drive the system towards solidification at the melting point, we have increased the density of the Fe liquid until we have observed nucleation of a solid phase. In order to achieve this on the short time scale accessible for our 3d simulations, we had to use enormous densities: $n_0 \geq 0.515$, which are evidently out of the density range accessible experimentally and of the validity range of our approximations. Accordingly, the present results need to be taken with reservations. Our findings are summarized in figures 6 and 7. For $n_0 \geq 0.5125$, we have seen first the nucleation of an amorphous solid phase (see figures 6a–d), which then soon transformed first into a bcc polycrystal, and later into a bcc single crystal (see figures 6e–h). The time of appearance for amorphous and crystalline phases is shown as a function of initial liquid density in figure 7. Remarkably, we have been unable to detect any phase transition for more than $10^6$ time steps for $n_0 = 0.51$, while slightly further, at $n_0 = 0.5125$, the amorphous phase have appeared in $\sim 2525$ time steps. These results suggest that crystal nucleation is clearly enhanced by the presence of the amorphous precursor phase, and direct bcc nucleation from the liquid phase requires several orders of magnitude longer time than via the precursor. Although the presence of an amorphous precursor phase has been reported in 2d colloidal crystallization experiments [16−18], and it has been predicted theoretically for simple liquids [29, 30], and amorphous precursors seem to be quite general in biological crystallization processes [106−110], we are unaware of any evidence supporting the presence of this behaviour in metallic systems. This lack of supporting evidence and the extreme conditions we have used warrant further investigations of freezing at ambient pressure combined with large undercoolings. Work is underway in this direction.

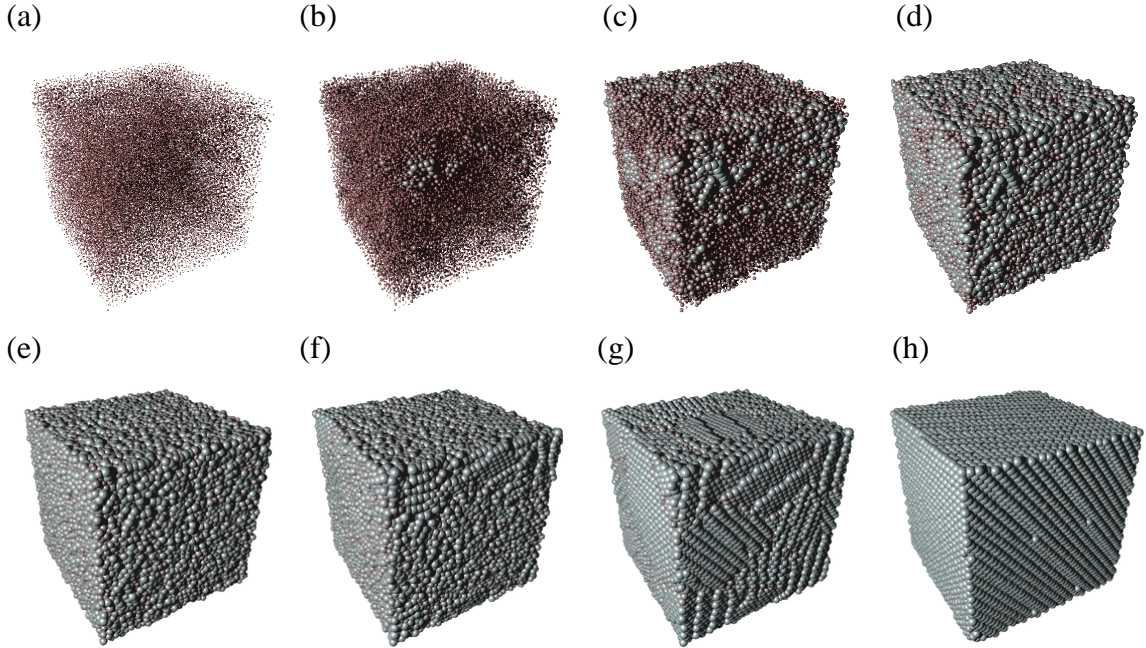

Figure 6. Snapshots of two-stage crystallization of highly compressed Fe melt ($n_0 = 0.54$) at the normal pressure melting temperature as predicted by the PFC EOF model: (a) − (d) transformation to amorphous solid (the images correspond to time steps 325, 350, 375 and 400); (d) − (h) nucleation and growth on the bcc phase (the images correspond to time steps 500, 1000, 2000 and 3500.) The simulation has been performed on a rectangular grid of size $300 \times 300 \times 300$. The localized particles are represented by spheres drawn around the respective density peaks with a diameter proportional to the amplitude of the density peak relative to the initial liquid density.

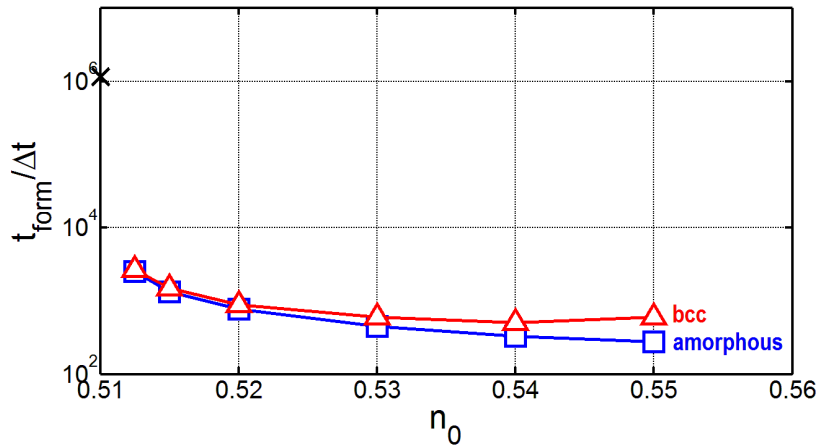

Figure 7. The time of appearance of the amorphous phase (squares) and the bcc crystal (triangles) in highly compressed Fe liquids as a function of the initial liquid density ($n_0$) as predicted by the PFC EOF model. The cross at $n_0 = 0.51$ indicates the time, at which no solidification has been yet observed.

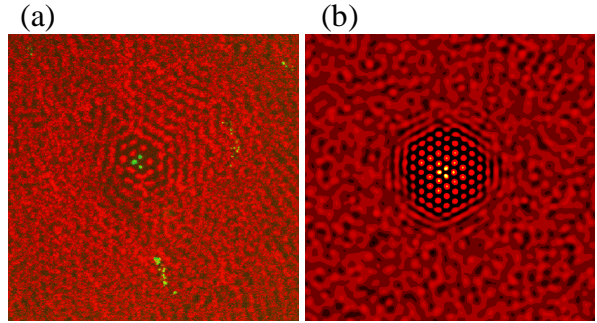

Figure 8. Patterning in experiment (left) vs. PFC simulation (right): (a) fluorescence microscopy image of time averaged ordering of particle density around particles of position *fixed by optical tweezers* (three green points at the centre) [111] © 2003 The Royal Society of Chemistry; and (b) small cluster initiated by a triplet of particles (yellow) held in position by a triangular triple-well potential in the PFC. ($V_1 = 0.5$, $r^* = -0.75$, $\alpha^* = 0.01$, and $\psi_0 = -0.5$; 2048 × 2048 grid. Reduced particle density map is shown.) Note the striking similarity of the experimental and simulation images.

### 3.4. *Modelling of liquid ordering, heteroepitaxy and heterogeneous nucleation in 2d*

The introduction of a potential energy term into the free energy density representing the patterning forces, such as those exerted by optical tweezers, leads to specific ordering of the particles in the PFC.

*A. Liquid ordering around fixed particles.* First, we model liquid ordering around particles held in fixed positions (e.g., by optical tweezers). In the simulation three particles are fixed into positions forming a triangle by a suitable potential. The ordering of the liquid the PFC predicts around the fixed particles, extends to several particle diameters, and resembles closely to images from fluorescence microscopy [111] (figure 8).

*B. Heteroepitaxy.* The effect of a crystalline wall on crystal growth can also be studied in the present model. For example, using the potential $V = V_1 [\cos(qx) + \cos(qy)]$ in a stripe across the centre of the simulation window, where $q = 2\pi/a_0$, and $a_0$ is the lattice constant of the square lattice, and adding an appropriate excess number density to the same region, after a short transient period particles aligned on a square lattice appear, which represent a square lattice substrate. Its influence on crystallization depends on the lattice constant (figure 9). If $a_0$ matches to inter-particle distances of one or other faces of the crystal, barrierless or low-barrier epitaxial growth takes place, and the choice of $a_0$ can be used to control the orientation of the forming crystal (*cf.* figures 9b and 9d). Otherwise, the crystalline phase may appear via nucleation at the interface (see figures 9a and 9c). Remarkably, the mechanism of epitaxial growth depends on the particle density in the fluid phase. Below about $\psi_0$ $\approx -0.506$, the diffusion-controlled mode is observed ($x \propto \tau^{1/2}$, where $x$ is the distance from the surface of the substrate) as opposed to the diffusionless mode appearing above this limit ($x \propto \tau$). The presence of these two growth modes is well known in colloidal systems [112–117]. A detailed study of the transition between the diffusion-controlled and diffusionless mechanisms will be presented elsewhere [118]. The driving force dependence of the local order at fixed lattice constant of the substrate is shown in figure 10. Whether the barrierless growth, seen beyond $\psi_0 \approx -0.506$ in these simulations, could be associated with the surface spinodal predicted by Models B and C specified in [59, 60] needs further investigations. We also find that, at large driving forces of crystalline aggregation, a layered structure composed of alternating 2d hexagonal and square-structured layers forms to release the stress from crystallizing to a non-equilibrium density (see figures 10c and 10d). This raises the question whether the square-lattice may exist as a metastable phase on its own in the PFC model.

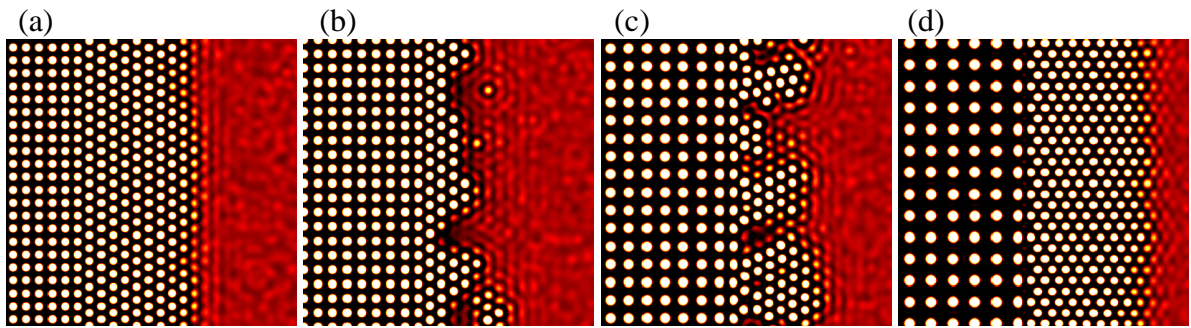

Figure 9. Heteroepitaxy in the PFC model on structured wall represented by periodic potential: (a) – (d) Growth on (01) interface of a square lattice: Effect of lattice constant of substrate ($a_0 = 18$, 20, 25 and 30 $\Delta x$ corresponding to $a_0/\sigma \approx 1.0$, 1.11, 1.39 and 1.67). Barrierless or low-barrier growth is observed if the (0 1) face of the substrate is commensurate to the (1 0 $\bar{1}$) type face of the 2d hexagonal structure. In panels (a) and (c) lattice mismatch prevents such immediate growth from the surface of the substrate; here crystallization takes place via nucleation and growth. (Fraction of the 50 $a_0$ × 50 $a_0$ simulation box is shown. Other simulation parameters were: $V_1 = 0.1$, $r^* = -0.75$, $\alpha^* = 0.01$, and $\psi_0 = -0.5$. Reduced particle density maps are shown.)

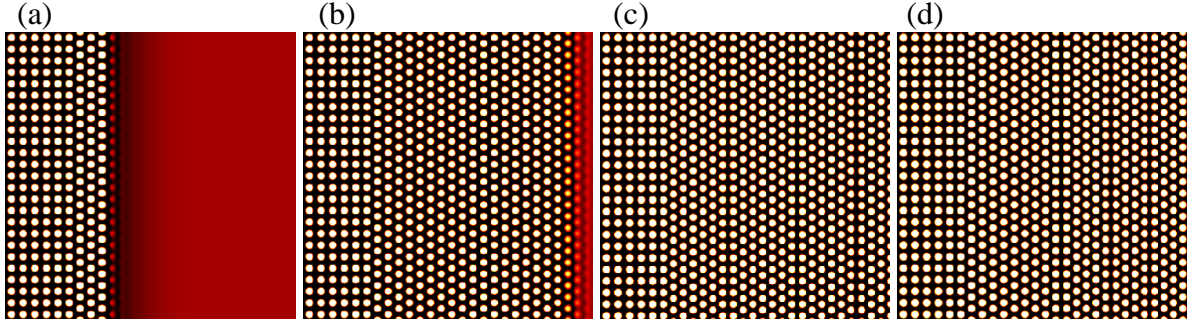

Figure 10. Snapshots of heteroepitaxial growth taken at a fixed reduced time ($\tau = 10\,000\Delta\tau$) in simulations performed under different supersaturations (the initial reduced particle density increases from left to right as $\psi_0 = -0.5075, -0.505, -0.47$ and $-0.45$). The upper right quarter of the $900 \times 900$ sized simulation boxes are shown. (Other simulation parameters are: $V_1 = 0.1$, $r^* = -0.75$, and $\alpha^* = 0$. Reduced particle density maps are shown.) Diffusion-controlled growth has been observed for panel (a), and a diffusionless growth of a lower density crystal for panels (b)–(d). Note the depletion zone ahead of the growth front in panel (a), the sharp change in growth velocity as a function of $\psi_0$, and that, with increasing velocity, the frequency of the square-lattice type stacking faults increases.

Our thermodynamic computations, based on the single-mode approximation to the particle density, indicate that indeed the square-lattice phase is metastable from the thermodynamic viewpoint. A metastable coexistence has been predicted between the liquid and square-lattice phases (for $r^* = -0.75$ the respective equilibrium liquid and solid densities are $\psi_L^e = -0.5394$ and $\psi_{Sq}^e = -0.2952$). However, its free energy density for the square-lattice is significantly higher than for the 2d hexagonal phase (figure 11). A further metastable coexistence has been predicted between the 2d hexagonal and the square-lattice phases (with equilibrium particle densities of $\psi_{Hex}^e = -0.1103$ and $\psi_{Sq}^e = 0.01930$), however, in a region, where the striped phase is the stable one. Our PFC simulations are in a general agreement with these predictions. In the presence of noise, the square-lattice seed crystallites (even the large ones) either melted or transformed into a faulty 2d hexagonal structure in the particle density range investigated ($\psi_0 \in [-0.55, -0.35]$). Remarkably, we have not seen the square-lattice to grow even in the absence of noise. Rather, if the square-lattice had not melted, the 2d hexagonal phase appeared on its surface, while the region occupied by the square-lattice has shrunk: The square-lattice transformed into faulty 2d hexagonal structure starting from its surface. This suggests that, in agreement with theoretical expectations for simple pair potentials [119], the square-lattice is probably mechanically unstable here.

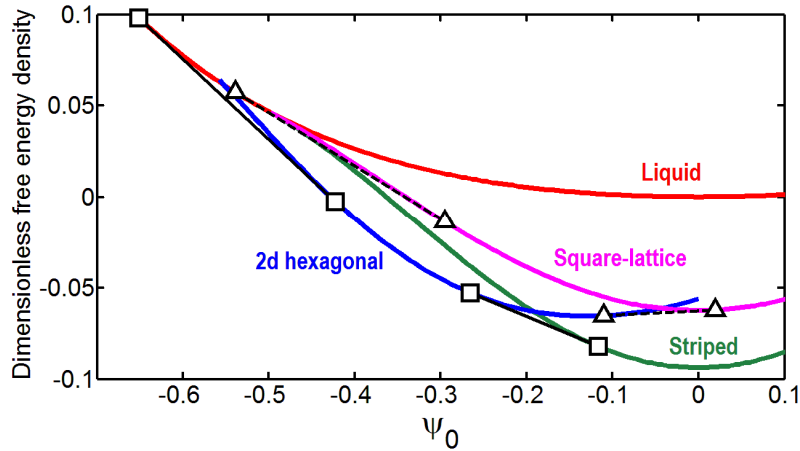

Figure 11. Thermodynamics of polymorphism in 2d in the PFC model at $r^* = -0.75$: Dimensionless free energy density vs. reduced particle density obtained using the single-mode approximation for the *stable* liquid, 2d hexagonal and striped phases and for the *metastable* square-lattice. (The stable coexistence regions are and the respective equilibrium densities are denoted by solid black lines and squares, while the dashed black lines and triangles stand for the metastable coexistences and the equilibrium densities of the respective phases.)

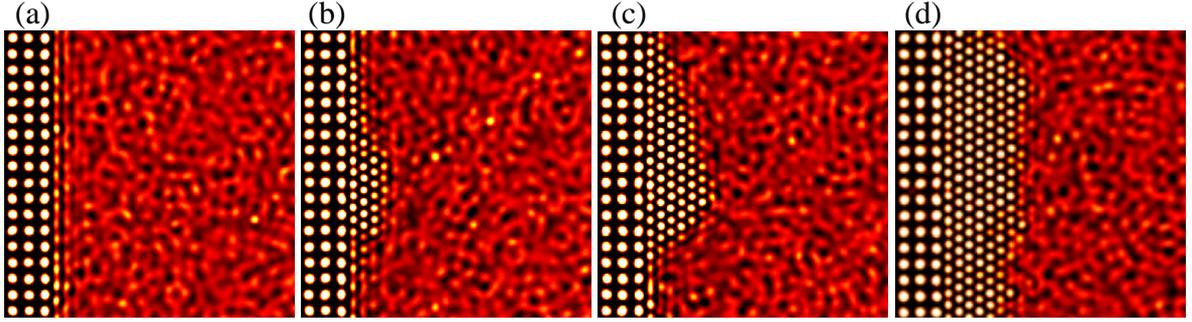

Figure 12. Snapshots of heterogeneous nucleation and growth on the (0 1) surface of a flat substrate of square-lattice ($a_0/\sigma \approx 1.39$): (a) – (d) Snapshots taken at dimensionless time steps 1000, 35 000, 80 0000, and 310 000. The upper right quarter of the 900 × 900 sized simulation box is shown. (Other simulation parameters were: $V_1 = 0.1$, $r^* = -0.25$, $\alpha^* = 0.1$, and $\psi_0 = -0.32$. Reduced particle density maps are shown.)

*C. Heterogeneous nucleation.* First, we study crystal nucleation on a flat surface of a square-lattice substrate. Here we use $r^* = -0.25$ corresponding to the relatively small anisotropy of metallic systems as pointed out above, and $a_0/\sigma \approx 1.39$, which provides sufficient mismatch to prevent immediate growth from the surface of the substrate. A sequence of snapshots displaying heterogeneous crystal nucleation and late stage growth morphology is presented in figure 12. A remarkable feature of the simulation is the large amplitude of the capillary fluctuations and the frequent appearance / disappearance of small crystalline clusters during the initial stage of solidification.

Next, we investigate the classical prediction that corners should be favourable places for crystal nucleation [50, 51]. This prediction is also shared by more advanced models, such as the coarse-grained continuum models of heterogeneous nucleation (phase-field and Cahn-Hilliard type models) [55–61] that cannot consider the atomic structure of the crystal. In our study, we assume a rectangular inner corner, while we investigate three possible structures for the substrate: (i) square-lattice with the (1 0) type faces parallel with the surface of the substrate; (ii) a square lattice, however, now rotated by 45° relative to the previous case; and (iii) a unstructured substrate represented by a repulsive (positive) value of the external potential inside the substrate. The results obtained at early and later stages of crystallization are shown in figure 13. Contrary to the classical expectation, *these* corners are *not favourable sites* for the nucleation of *this* crystal. The reason is fairly clear: whether the preferred orientation of the hexagonal crystal is that with a (1 0 $\bar{1}$) type face parallel or perpendicular to the surface of the substrate, the crystallites forming on the two perpendicular faces of the substrate have different crystallographic orientations, so when impinging upon each other they need to form a grain boundary. The same stands for the corner: due to the incompatibility of the symmetries of the crystal and the corner, the nucleus that would appear in the corner shall contain energetically costly defects. Evidently, a 60° corner of the unstructured substrate or a 2d hexagonal crystal structure of the substrate would remove this frustration and make the corner a favourable site for crystal nucleation.

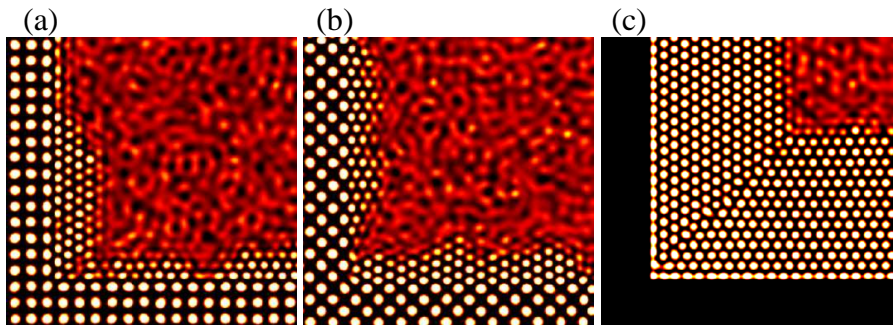

Figure 13. Heterogeneous nucleation in rectangular inner corners ($a_0/\sigma \approx 1.39$): (a) On (01) surfaces of a square lattice; (b) on (11) surfaces of a square lattice; and (c) on an unstructured substrate. (The upper right quarter of the 900 × 900 sized simulation boxes is shown. Other simulation parameters were: $r^* = -0.25$, $\alpha^* = 0.1$, $\psi_0 = -0.32$, and $V_1 = 0.1$ for the square-lattice substrates, while for the unstructured substrate we have prescribed $V_0 = 0.5$ inside the substrate. Reduced particle density maps are shown.) Note the frustration at the corner and the formation of a grain boundary starting from the corner at later stages.

### 3.5. *Modelling of colloid patterning in 2d*

In this section, we are going to address controlled self-assembly of colloid particles in the presence of modulated substrates, where the latter will be represented by appropriate potential energy terms in our PFC simulations.

First, we model colloid patterning under the influence of periodic substrates, which can be realized via creating patches that are chemically attractive to the colloidal particles [120]. Depending on the size of the patches single, double, triple, etc., occupations of the patches are possible (figure 14a), and depending on the distance of the patches various states can be realized, as predicted by Langevin simulations in which the patterned substrate is represented by appropriate periodic potentials [120]. A PFC model supplemented by periodic potential is suitable for such studies. Introducing circular potential wells arranged on a square lattice, and varying the diameter of the attractive wells as well as their distance, we have been able to reproduce the patterns seen in the experiments (see figure 14b).

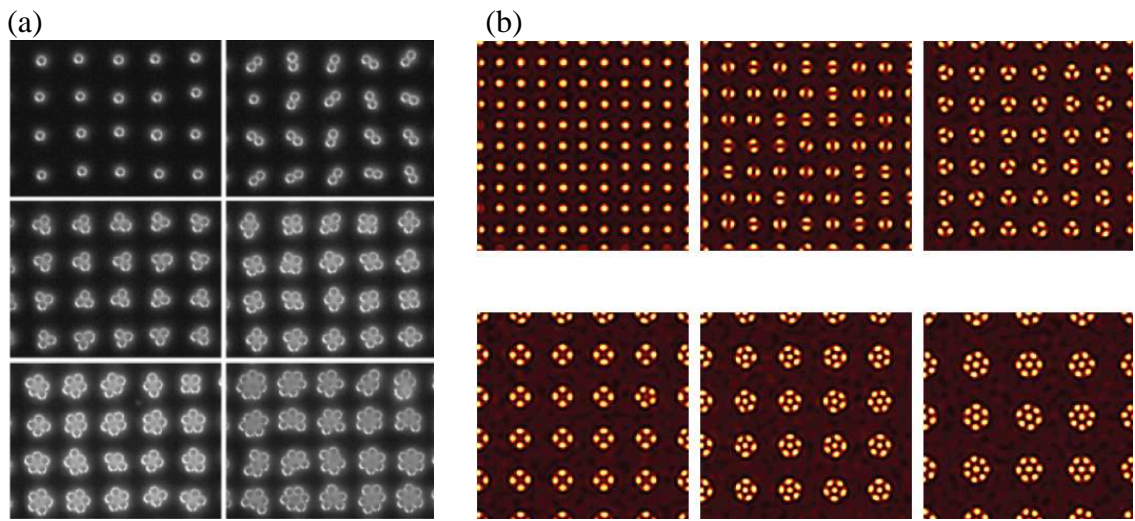

Figure 14. (a) Single and multiple occupation of chemically patterned periodic substrate by colloidal particles as a function of increasing patch size [120] © 2002 Wiley-WCH Verlag GmbH. (b) PFC simulations with increasing diameter of circular attractive potential wells (right). Fractions of $818 \times 818$ simulations are shown. Other simulation parameters were: depth of circular wells $V_0 = -0.5$, $r^* = -0.75$, $\alpha^* = 0.1$, and $\psi_0 = -0.75$. Reduced particle density maps are shown. The ratio of the potential well diameters relative to the single occupation case has been 1, 1.25, 1.5, 2, 2.13 and 2.5.

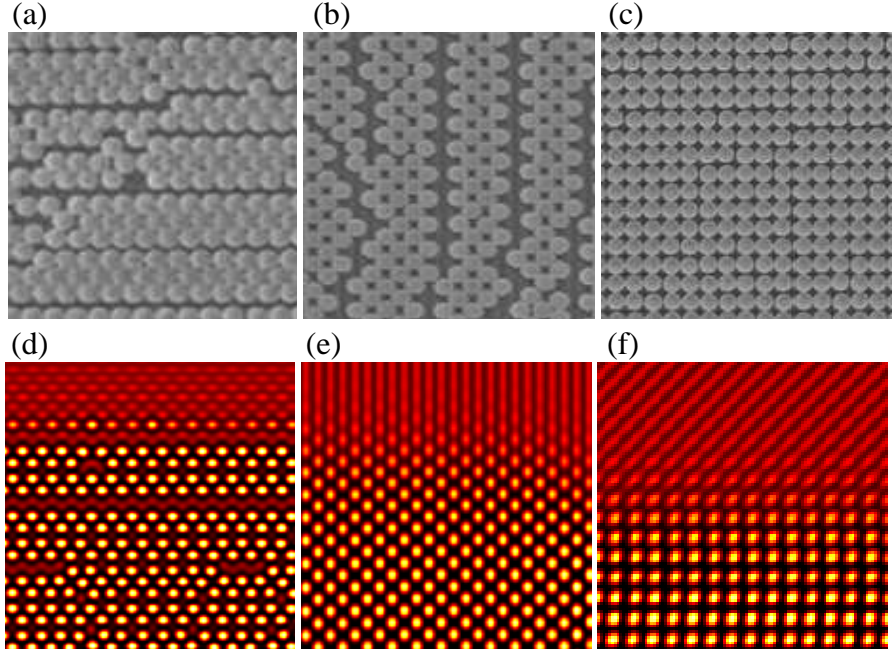

Figure 15. Fabrication of square colloidal crystals via controlled self-assembly in experiment and PFC simulations. (a) – (c): Optical microscopy image on the ordering of particles due to grooves tilted relative to the growth front (tilting angles: 0°, 90°, and 45°) [123] © 2008 Wiley-WCH Verlag GmbH & Co. KGaA; PFC simulations of (d) – (f) particle ordering in the presence of potential approximating the effect of combined forces from capillarity and grooved substrate underlying the liquid surface. (Fractions of $1100 \times 1100$ simulations are shown. Other simulation parameters were: $V_1 = 0.3$, $r^* = -0.75$, $\alpha^* = 0$, and $\psi_0 = -0.495$. Reduced particle density maps are shown.)

Next, we model colloidal self-assembly under the effect of capillary-immersion forces occurring when capillarity at the air-liquid interface and the varying immersion due to a grooved surface underlying the layer of colloid particles interact with each other. Experiments of this kind have been used to produce single and double particle chains [122] and the otherwise unfavorable square-lattice structure [123]. The resultant of the capillary immersion forces can be expressed as [122]

$$F = \gamma(2\pi r_c)\sin\psi_c \frac{(\eta q)\cos(qx)}{\sqrt{1+[(\eta q)\cos(qx)]^2}}, \qquad (16)$$

where $\gamma$ is the surface tension of the liquid-air interface, $r_c$ is the radius of the contact line on the spherical colloidal particle, $\psi_c$ is the is the mean slope angle of the meniscus at the contact line, $\eta$ is the amplitude of the surface undulations, $q = 2\pi/\lambda$, while $\lambda$ is the wavelength of the surface undulations (ripples/grooves). Under the conditions of the experiments [122], such forces can be well represented by a potential of the form $V = V_1 \cos(qx)$, where $V_1$ depends on $\gamma$, $\psi_c$, and $r_c$.

Setting $\lambda = \sigma/2^{1/2}$, where $\sigma$ is the inter-particle distance and varying the orientation of the grooves relative to the crystallization (drying) front, patterns seen in the experiments [123] are observed to form in the PFC simulations (figure 15): For grooves parallel to the front, a frustrated 2D hexagonal structure of randomly alternating double and triple layers, separated by channels appear. When the grooves are perpendicular to the crystallization front, the particles align themselves on a square lattice with the (1 1) orientation lying in the interface, while for a 45° declination of the grooves the same 2D square structure forms, however, now with the (1 0) face lying in the front. In these simulations, solidification has been started by increasing the local density of the liquid in a stripe at the centre of the simulation, which together with noise lead to the formation of two roughly planar crystallization fronts propagating into opposite directions.

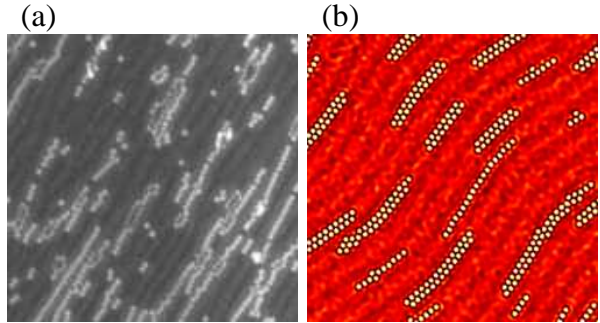

(a) (b)

Figure 16. Patterning in experiment (left) vs. PFC simulation (right): (a) single and double particle chains evolving due to capillary-immersion forces on the surface of a rippled substrate [122] © 2006 American Chemical Society; and (b) the particle chains forming in the PFC model using a tilted an wavy version of the potential described in the text. (Fraction of a 2400 × 1200 simulation is shown. Other simulation parameters were: $V_1 = 0.05$, $r^* = -0.75$, $\alpha^* = 0.1$, and $\psi_0 = -0.6$. Reduced particle density map is shown.)

In the case of a homogeneous initial particle density, we observe the nucleation and growth of single and double chains (figure 16), which in the case of wavy and tilted potential wells lead to morphologies resembling closely to the experiments [122].

## 4. Conclusions

We have used a simple dynamical density functional theory to address microscopic aspects of various solidification problems including homogeneous and heterogeneous nucleation, heteroepitaxy, and self-organized pattern formation in 2d colloidal systems in the presence of capillary immersion forces due to modulated substrate. Our PFC simulations have shown the following.

(i) In agreement with recent experiments on 2d colloidal systems [15], at small/medium supersaturations crystallization takes place via direct homogeneous nucleation of the 2d hexagonal phase and its growth, without the appearance of the hexatic phase.

(ii) At high supersaturations, 2d crystallization happens via the appearance of an amorphous precursor phase. This finding accords with recent experiments on 2d colloidal systems [16–18].

(iii) Bcc crystal nucleation from extremely compressed Fe liquids happens via an amorphous precursor phase. On the time-scale of our simulations no direct bcc nucleation has occurred from the liquid phase.

(iv) Contrary to expectations from the classical nucleation theory, corners are not necessarily favourable places of heterogeneous crystal nucleation: The interplay of the structure of the forming crystal, and of the structure and geometry of the substrate decides whether the corner helps or suppresses crystal nucleation.

(v) Supplementing the PFC with appropriate potential terms represents a powerful tool to model the dynamics of self-organized pattern formation in the presence of modulated substrates.

## Acknowledgements

We thank Mathis Plapp (École Polytechnique, CNRS, Palaiseau, France) for the enlightening discussions on noise and for the critical reading of our manuscript, and to Péter Hartmann (Research Institute for Solid State Physics and Optics, Budapest, Hungary) for the discussions on 2d melting/crystallization and for providing us the code, we used in evaluating the pair correlation and bond-order correlation functions. This work has been supported by the EU FP7 Collaborative Project ENSEMBLE under Grant Agreement NMP4-SL-2008-213669 and by the Hungarian Academy of Sciences under contract OTKA-K-62588. T. P. has been supported by the Bolyai János Scholarship.

# References

[1] J. Kosterlitz and D. Thouless, J. Phys. C **6** 1181 (1973); B. Halperin and D. Nelson, Phys. Rev. Lett. **41** 121 (1978); A. Young, Phys. Rev. B **19** 1855 (1979).
[2] For a recent review see H. H. Grünberg, P. Keim, and G. Maret, in *Soft Matter, Vol. 3: Colloidal Order from Entropic and Surface Forces.* Edited by G. Gompper and M. Schick (WILEY-VCH Verlag GmbH & Co. KGaA, Weinheim, 2007) pp. 40-83.
[3] K. Binder, S. Sengupta and P. Niebala, J. Phys.: Condens. Matter **14** 2323 (2002).
[4] A. Jaster, Phys. Lett, A **330** 120 (2004).
[5] X. Qi, Y. Chen, Y. Jin, *et al.* J. Korean Phys. Soc. **49** 1682 (2006).
[6] S. Z. Lin, B. Zheng and S. Trimper, Phys. Rev. E **73** 066106 (2006).
[7] C. A. Murray and D. H. Van Winkle, Phys. Rev. Lett. **58** 1200 (1987).
[8] Y. Tang, A. J. Armstrong, R. C. Mockler, *et al.*, Phys. Rev. Lett. **62** 2401 (1989).
[9] R. Kusner, J. Mann, J., Kerins, *et al.*, Phys. Rev. Lett. **73** 3113 (1994).
[10] A. Marcus and S. Rice, Phys. Rev. Lett. **77** 2577 (1996).
[11] P. Keim, G. Maret and H. H. von Grünberg, Phys. Rev. E **75** 031402 (2007).
[12] C. H. Mak, Phys. Rev. E **73** 065104 (2006).
[13] K. Chen, T. Kaplan and M. Mostoller, Phys. Rev. Lett. **74** 4019 (1995).
[14] F. L. Somer, G.S. Canright, T. Kaplan, *et al.*, Phys. Rev. Lett. **79** 3431 (1997).
[15] P. Dillmann, G. Maret and P. Keim, J. Phys.: Condens. Matter **20** 404216 (2008).
[16] T. J. Zhang and X. Y. Liu, J. Am. Chem. Soc. **129**, 13520 (2007).
[17] J. R. Savage and A. D. Dinsmore, Phys. Rev. Lett. **102** 198302 (2009).
[18] J. J. Yoreo, to be published.
[19] S. Auer and D. Frenkel, Nature **409** 1020 (2001).
[20] D. Moroni, P. R. ten Wolde and P. G. Bolhuis, Phys. Rev. Lett. **94** 235703 (2005).
[21] P. R. ten Wolde, M. J. Ruiz-Montero and D. Frenkel, Phys. Rev. Lett. **75** 2714 (1995).
[22] J. Hernández-Guzmám and E. R. Weeks, Proc. Nat. Acad. Sci. **106** 15198 (2009).
[23] P. R. ten Wolde and D. Frenkel, Phys. Chem. Chem. Phys. **1** 2191 (1999).
[24] S. Alexander and J. McTague, Phys. Rev. Lett. **41** 702 (1978).
[25] B. Groh and B. Mulder, Phys. Rev. B **59** 5613 (1999).
[26] W. Klein, Phys. Rev. E **64** 056110 (2001).
[27] P. R. ten Wolde, M. J. Ruiz-Montero and D. Frenkel, J. Chem. Phys. **104** 9932 (1996).
[28] T. A. Johnson and C. Elbaum, Phys. Rev. E **62** 975 (2000).
[29] J. F. Lutsko and G. Nicolis, Phys. Rev. Lett. **96** 046102 (2006).
[30] J. Berry, K. R. Elder and M. Grant, Phys. Rev. E **77** 06506 (2008).
[31] P. R. ten Wolde and D. Frenkel, Science **277** 1975 (1997).
[32] V. Talanquer and D. W. Oxtoby, J. Chem. Phys. **109** 223 (1998).
[33] R. P. Sear, J. Chem. Phys. **114** 3170 (2001).
[34] A. Shiryayev and J. D. Gunton, J. Chem. Phys. **120** 8318 (2004).
[35] G. I. Tóth and L. Gránásy, J. Chem. Phys. **127** 074710 (2007).
[36] S. Auer and D. Frenkel, Phys. Rev. Lett. **91** 015703 (2003).
[37] K. Yasuoka, G. T. Gao, and X. C. Zeng, J. Chem. Phys. **112** 4279 (2000).
[38] A. Milchev, A. Milchev, and K. Binder, Comp. Phys. Comm. **146** 38 (2002).
[39] E. B. Webb III, G. S. Grest, and D. R. Heine, Phys. Rev. Lett. **91** 236102 (2003).
[40] A. Esztermann and H. Löwen, J. Phys.: Condens. Matter **17** S429 (2005).
[41] See. e.g.: W. Kurz and D. J. Fisher, *Fundamentals of Solidification.* (Trans Tech Publ., Switzerland, 1989), Chap. 2; K. F. Kelton, Solid State Phys. **45** 75 (1991).
[42] J. Aizenberg, A. J. Black and G. M. Whitesides, Nature **398** 495 (1999).
[43] F. Favier, E. C. Walter, M. P. Zach, *et al.*, Science **293** 2227 (2001).
[44] J. D. Hartgelink, E. Beniash, and S. I. Stupp, Science **294** 1684 (2001).
[45] N. Ravishankar, V. J. Shenoy, and C. B. Carter, Adv. Mater. **16** 76 (2004).
[46] C. Y. Nam, J. Y. Kim and J. E. Fischer, Appl. Phys. Lett. **86** 193112 (2005).
[47] A. L. Briseno, S. C. B. Mannsfeld, M. M. Ling, *et al.*, Nature **444** 913 (2006).
[48] For example: C. Herring, Chap. 8 of *The Physics of Powder Metallurgy*, ed. W. E. Kingston (McGrew-Hill, New York, 1951), p. 143.
[49] M. Volmer, Z. Elektrochem. **35** 555 (1929).
[50] D. Turnbull. J. Chem. Phys. **18** 198 (1950).
[51] See e.g., J. W. Christian, *Transformations in Metals and Alloys.* (Pergamon, Oxford, 1981).

[52]   A. L. Greer, A. M. Bunn, A. Tronche, *et al.*, Acta Mater. **48** 2823 (2000).
[53]   T. D. Quested and A. L. Greer, Acta Mater. **53** 2683 (2005).
[54]   S. A. Reavley and A. L. Greer, Philos. Mag. **88** 561 (2008).
[55]   M. Castro, Phys. Rev. B **67** 035412 (2003).
[56]   L. Gránásy, T. Pusztai, T. Börzsönyi, *et al.*, J. Mater. Res. **21** 309 (2006).
[57]   H. Emmerich, R. Siquieri, J. Phys.: Cond. Matter **18**, 11121 (2006).
[58]   R. Siquieri, H. Emmerich, Philos. Mag. Lett. **87**, 829 (2007).
[59]   L. Gránásy, T. Pusztai, D. Saylor, *et al.*, Phys. Rev. Lett. **98** 035703 (2007).
[60]   J. A. Warren, T. Pusztai, L. Környei, *et al.*, Phys. Rev. B. **79** 014204 (2009).
[61]   J. J. Hoyt and L. N. Bush, J.Chem. Phys. **78** 1589 (1995).
[62]   M. Schrader, P. Virnau, D. Winter, *et al.*, Eur. Phys. J. Special Topics **177** 103 (2009).
[63]   S. van Teeffelen, C. N. Likos and H. Löwen, Phys. Rev. Lett. **100** 108302 (2008).
[64]   K. R. Elder, M. Katakowski, M. Haataja, et al., Phys. Rev. Lett. **88** 245701 (2002).
[65]   R. Prieler, J. Hubert, D. Li, *et al.*, J. Phys.: Condens. Matter **21** 464110 (2009).
[66]   C. V. Achim, M. Karttunen, K. R. Elder, *et al.*, Phys. Rev. E **74** 021104 (2006).
[67]   A. Jaatinen, C. V. Achim, K. R. Elder, *et al.*, Phys. Rev. E. **80** 031602 (2009).
[68]   K. R. Elder and M. Grant, Phys. Rev. E **70** 051605 (2004),
[69]   T. V. Ramakrishnan and M. Yussouff, Phys. Rev. B **19** 2775 (1979).
[70]   S. A. Brazovskii, Zh. Eksp. Teor. Fiz. **68** 175 (1975).
[71]   J. Swift and P. C. Hohenberg, Phys. Rev. A **15** 319 (1977).
[72]   S. van Teeffelen, R. Backofen, A. Voigt, *et al.*, Phys. Rev. **79** 051404 (2009).
[73]   K. A. Wu and A. Karma, Phys. Rev. B **76** 184107 (2007).
[74]   S. Majaniemi and N. Provatas, Phys. Rev. E **79** 011607 (2009).
[75]   G. Tegze, L. Gránásy, G. I. Tóth, *et al.*, Phys. Rev. Lett. **103** 035702 (2009).
[76]   K. R. Elder, N. Provatas, J. Berry, *et al.*, Phys. Rev. B **75** 064107 (2007).
[77]   N. Provatas, J. A. Dantzig, B. Athreya, *et al.*, JOM **59** 83 (2007).
[78]   T. Pusztai, G. Tegze, G. I. Tóth, *et al.*, J. Phys.: Condens. Matter **20** 404205 (2008).
[79]   G. Tegze L. Gránásy, G. I. Tóth, *et al.*, J. Comput. Phys. **228**, 1612 (2009).
[80]   J. Berry, K. R. Elder and M. Grant, Phys. Rev. B **77** 224114 (2008).
[81]   J. Mellenthin, A. Karma and M. Plapp, Phys. Rev. B **78** 184110 (2008).
[82]   S. Majaniemi and M. Grant, Phys. Rev. B **75** 054301 (2007); S. Majaniemi, personal communication (2009).
[83]   M. Plapp cited in A. Karma and W.J. Rappel, Phys. Rev. E **60** 3614 (1999).
[84]   M. Frigo and S.G. Johnson, Proc. IEEE **93** 216 (2005).
[85]   U. M. B. Marconi and P. Tarazona, J. Chem. Phys. **110** 8032 (1999).
[86]   H. Löwen, J. Phys.: Condens. Matter **15** V1 (2003).
[87]   A. Archer and M. Rauscher, J. Phys. A **37**, 9325 (2004).
[88]   A. Karma, personal communication.
[89]   See application for the Brazowskii/Swift-Hohenberg model in N. A. Gross, M. Ignatiev and B. Chakraborty, Phys. Rev. E **62** 6116 (2000).
[90]   M. Plapp, this volume.
[91]   L. Leibler, Macromol. **13** 1602 (1980).
[92]   F. S. Bates and G. H. Fredrickson, Physics Today **52** 32 (1999).
[93]   V.P. Skripov, in Crystal Growth and Materials, eds. E. Kaldis and H.J. Scheel (North Holland, Amsterdam, 1976) pp. 327.
[94]   D. W. Oxtoby, Annu. Rev. Mater. Res. **32** 32 (2002).
[95]   L. S. Bartell and D. T. Wu, J. Chem. Phys. **127** 174507 (2007).
[96]   M. Elenius and M. Dzugutov, J. Chem. Phys. **131** 104502 (2009).
[97]   J. D. Gunton, M. San Miguel and P. Sahni, P., in Phase Transitions and Critical Phenomena, Vol. 8, eds. C. Domb and J. L. Lebowitz (Academic, London, 1983) pp. 267–466.
[98]   R. Backofen and A. Voigt, J. Phys.: Condens. Matter. **21** 464109 (2009).
[99]   V. A. Shneidman and R. K. P. Zia, Phys. Rev. E **63**, 085410 (2001).
[100]  N. Akutsu and Y. Akutsu, J. Phys. Soc. Jpn. **56** 2248 (1987).
[101]  G. Y. Onoda, Phys. Rev. Lett. **55** 226 (1985).
[102]  A. T. Skjeltorp, Phys. Rev. Lett. **58** 1444 (1987).
[103]  R. K. Kalia and P. Vashishta, J. Phys. C: Solid State Phys. **14** L643 (1981).
[104]  P. Ballone, G. Pastore, M. Rovere, *et al.* J. Phys. C: Solid State Phys. **18** 4011 (1985).
[105]  A. D. Law and D. M. A. Buzza, J. Chem. Phys. **131** 094704 (2009).
[106]  I. M. Weiss, N. Tuross, L. Addadi, *et al.*, J. Exp. Zool. **293** 478 (2002).
[107]  L. Addadi, S. Raz, S. Weiner, Adv. Mater. **15** 959 (2003).

[108] Y. Politi, T. Arad, E. Klein, *et al.*, Science **306** 1161 (2004).
[109] J. Mahamid, A. Sharir, L. Addadi, *et al.*, Proc. Natl. Acad, Sci. U. S. A. **105** 12748 (2008).
[110] J. Tao, H. Pan, H. Zhai, *et al.*, Cryst. Growth and Design **9** 3154 (2008).
[111] A. van Blaaderen, J. P. Hoogenboom, D. L. J. Vossen, *et al.*, Faraday Discuss. **123** 107 (2003).
[112] A. P. Gast and Y. Monovoukas, Nature **351** 553 (1991).
[113] K. Schätzel and B. J. Ackerson, Phys. Rev. E **48** 3766 (1993).
[114] B. Russel, P. M. Chaikin, J. Zhu, *et al.*, Langmuir **13** 3871 (1997).
[115] S. Derber, T. Palberg, K. Schatzel, *et al.*, Physica A **235** 204 (1997).
[116] S. I. Henderson and W. van Megen, Phys. Rev. Lett. **80** 877 (1998).
[117] R. Wild and P. Harrowell, J. Chem. Phys. **114** 9059 (2001).
[118] G. Tegze, L. Gránásy, G. I. Tóth, *et al.*, to be published.
[119] For example: K. Zabinska, Phys. Rev. B **43** 3450 (1991).
[120] I. Lee, H. Zheng, M. F. Rubner, *et al.*, Adv. Mater. **14** 572 (2002).
[121] C. Reichhard and C. J. Olson, Phys. Rev. Lett. **88** 248310 (2002).
[122] A. Mathur, A-D. Brown, and J. Erlebacher, Langmuir **22** 582 (2006).
[123] J. Sun, Y. Li, H. Dong, *et al.*, Adv. Mater. **20** 123 (2008).